\pdfoutput=1
\documentclass[10pt,conference]{IEEEtran}
\usepackage[framemethod=tikz]{mdframed}
\let\labelindent\relax
\usepackage{lipsum}
\usetikzlibrary{shadows}
\usepackage{mathtools}
 \usepackage{paralist}
\usepackage{graphics}
\newmdenv[
tikzsetting= {fill=gray!10},
linewidth=1pt,
roundcorner=2pt, 
shadow=false
]{myshadowbox}
\usepackage{colortbl} 
\usepackage{subcaption}
\usepackage{blindtext, graphicx}
\usepackage{textcomp}
\usepackage{mathtools}
\usepackage{hyperref}
\usepackage{amsmath}
\usepackage{amsfonts}
\usepackage{caption}
\usepackage{color}
\usepackage[final]{listings}
\usepackage{graphicx} 
\usepackage{multirow}
\usepackage{balance}
\usepackage{picture}
\usepackage{relsize}
\usepackage{soul}
\usepackage{enumitem}
\setitemize{noitemsep,topsep=0pt,parsep=0pt,partopsep=0pt}
\usepackage{array}
\usepackage{makecell}
\definecolor{light-gray}{gray}{0.80}

\usepackage{verbatim}
\usepackage{algorithm}
\usepackage{algorithmicx}
\usepackage{algpseudocode}
\usepackage[export]{adjustbox}
\usepackage{mathtools}
\renewcommand{\footnotesize}{\scriptsize}
\definecolor{lightgray}{gray}{0.8}
\definecolor{darkgray}{gray}{0.6}

\definecolor{Gray}{rgb}{0.88,1,1}
\definecolor{Gray}{gray}{0.85}
\definecolor{Blue}{RGB}{0,29,193}
\usepackage{booktabs}
\usepackage{xspace}
\usepackage{todonotes}
\newcommand{\cusconfig}{\includegraphics[scale=0.5]{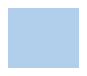}}
\newcommand{\cusmonrp}{\includegraphics[scale=0.5]{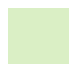}}
\newcommand{\cuspom}{\includegraphics[scale=0.5]{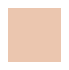}}
\newcommand{\cusxomo}{\includegraphics[scale=0.5]{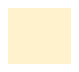}}

\definecolor{MyDarkBlue}{rgb}{0,0.08,0.45} 
\usepackage{array}
\newcolumntype{L}[1]{>{\raggedright\let\newline\\\arraybackslash\hspace{0pt}}m{#1}}

\usepackage{fancyvrb}
\fvset{%
fontsize=\small,
numbers=left
}

\definecolor{lightgray}{gray}{0.7}
\tikzstyle{highlighter} = [
  lightgray,
  line width = \baselineskip,
]

\newcounter{highlight}[page]

\AtBeginShipout{\AtBeginShipoutUpperLeft{\ifthenelse{\value{highlight} > 0}{\tikz[remember picture, overlay]{\foreach \stroke in {1,...,\arabic{highlight}} \draw[highlighter] (begin highlight \stroke) -- (end highlight \stroke);}}{}}}

\usepackage{soul}
\usepackage{color}

\definecolor{awesome}{rgb}{1.0, 0.13, 0.32}

\usepackage{amsmath}
\definecolor{Gray}{gray}{0.95}
\definecolor{LightGray}{gray}{0.975}

\newcommand{\bi}{\begin{compactitem}}%[leftmargin=0.4cm]}
\newcommand{\ei}{\end{compactitem}}
\newcommand{\be}{\begin{enumerate}}
\newcommand{\ee}{\end{enumerate}}

\usepackage{siunitx}

\definecolor{arsenic}{rgb}{0.23, 0.27, 0.29}

\begin{document}

\pagestyle{plain}

\title{FLASH: A Faster Optimizer for SBSE Tasks}

\author{\IEEEauthorblockN{Vivek Nair}
\IEEEauthorblockA{Dept. of Computer Science\\
North Carolina State University,\\
Raleigh, NC, USA\\
vivekaxl@gmail.com}
\and
\IEEEauthorblockN{Zhe Yu}
\IEEEauthorblockA{Dept. of Computer Science\\
North Carolina State University,\\
Raleigh, NC, USA\\
zyu9@ncsu.edu}
\and
\IEEEauthorblockN{Tim Menzies}
\IEEEauthorblockA{Dept. of Computer Science\\
North Carolina State University,\\
Raleigh, NC, USA\\
tim.menzies@gmail.com}
}

\maketitle

% human readabiloity
% - the  "one slide" rule

% TRANSPARENCY
% High readability is one of the advantages of decision trees.
% The relationship between the recommended information and
% user preferences is clear. Furthermore, the structure of the
% decision tree itself represents user preferences, so detailed
% information can be offered. 
% Movie Recommender System Based

% Mnimumim message length:
% size of the model/errors

% microsoft sumemr 2011 2012.
% nagapapan decisions for userd

% suppport for subequent plan genration

\begin{abstract}

Most problems in search-based software engineering involves balancing conflicting objectives. Prior  approaches to this task have 
required a large number of evaluations-- making them very slow to
execute and very hard to comprehend.
To solve these problems, this paper introduces FLASH, a decision-tree based optimizer that incrementally grows one decision
tree per objective. These
trees are then  used to select the next sample.  

This paper compares  FLASH 
  to state-of-the-art algorithms from  search-based SE and machine learning. This comparison
  uses   multiple SBSE  case studies  for 
release planning, configuration control, process modeling, 
and sprint planning for agile development. 
FLASH was found to be the fastest optimizer
(sometimes requiring less than 1\% of the evaluations
used by evolutionary algorithms).
Also, measured in terms of model size,
FLASH's reasoning was  far more succinct and comprehensible.
Further, measured in terms of finding effective optimization,
FLASH's recommendations were highly competitive with  other approaches. Finally, FLASH scaled to more complex models since it always terminated (while  state-of-the-art algorithm did not).

\end{abstract}

%\keywords{
%Performance Prediction, SBSE, Sampling, Rank-based method}

\IEEEpeerreviewmaketitle

\section{Introduction}\label{sec:intro}

\noindent
In 2001, Harman and Jones characterized a wide range 
of SE tasks as a optimization  problem; i.e 
\begin{quote}
{\em One in which 
optimal or near optimal solutions are sought in a search space of candidate solutions guided by a fitness function that distinguishes between and worse searches.}~\cite{harman2012search}.
\end{quote}
To complete such tasks, some algorithm must navigate
through a complex space of constraints. Often ``trade-offs''
must be made between multiple competing objectives; e.g.
\begin{enumerate}
\item[(a)]
 An optimized agile project selects use cases from a backlog    that deliver the {\em most} requirements in the {\em least} time~\cite{port08};
 \item[(b)]
 An optimized waterfall  project uses   {\em fewest} programmers to deliver  {\em most} code with  {\em fewest} bugs in 
 {\em least} time~\cite{menzies2009avoid};
 \item[(c)]
 An optimized release planner delivers the {\em most} functionality with the {\em least} cost while
 {\em minimizing} risks and {\em  maximizing} customer satisfaction~\cite{zhang2007moea};
 \item[(d)]
 An optimized configuration engine {\em most} explores the measure of interest of the  stakeholder (such as throughput) after reflecting on the {\em fewest} configurations~\cite{siegmund2012predicting}.
\end{enumerate}
A perennial question is which algorithm to use to optimize these tasks. Recent literature reviews report that the  search-based SE (SBSE) community makes extensive use of only two or three evolutionary algorithms~\cite{chen2017beyond}. 
This is curious since:
\begin{enumerate}
\item[(1)]
Researchers in software analytics have shown
that their standard algorithms can be readily adapted
to optimizing SE tasks -- sometimes significantly out-performing  EA methods~\cite{krall2015gale,nair17,Nair2016}.
\item[(2)]
For such optimization tasks, machine learning researchers prefer   non-evolutionary algorithms called ``Bayesian optimization''~\cite{bpo1,bpo2,zuluaga2013active, zuluaga2016varepsilon}.
\item[(3)]
Researchers in evolutionary algorithms (EAs)
are always improving their  algorithms~\cite{deb2002fast,deb2014evolutionary,zhang2007moea}.
\end{enumerate}
Given  plethora of methods 
there are two natural next questions.
Firstly, {\em should the SBSE
community consider changing their preferred set of algorithms? }
Secondly,
{\em is there any advantage in combining ideas from different research communities?}

To answer these questions,
this paper  experiments with a new optimizer called FLASH that combines ideas from the research communities (1), (2), (3) listed above.  When tested on SE tasks (a), (b), (c), (d), FLASH was often a ``better''
 optimizer than any of the standard methods.
 Here, by ``better'' we mean FLASH is far faster than the other algorithms studied in this paper. Also, FLASH  produces simpler and more comprehensible summaries of its conclusions. Further, FLASH's  proposed optimizations are often just as good as the other algorithms. 
Lastly, FLASH  scales to  more  complex models  since, it is the only algorithm in this particular case study that successfully terminated
on the   (a), (b), (c), (d) tasks.

Accordingly, we make several conclusions.
Firstly,  SBSE researchers
have been unnecessarily limiting themselves due to their algorithm selection. Secondly,
better SBSE techniques are now available based on a combination of methods from other fields.
As to the  specific  technical contributions of this work, the unique features of this paper are:
\bi
\item A demonstration that EAs may not be the preferred approach for some SBSE tasks;
\item A proof-of-concept algorithm, FLASH, that demonstrates
the value of combining insights and algorithms from
several fields;
\item An extension of standard SE configuration optimization to multi-objective reasoning;
\item A set of experiments showing the value of FLASH;
\item A public-domain version of FLASH and public-domain executable  versions of models used in (a), (b), (c), (d)\footnote{http://github.com/blinded4review}.
\item Guidelines  for when not to use FLASH.
\ei
% Further to the last point-- our experiments highlight not only the advantages of FLASH,
% but also  a class of models where we would recommend other algorithms. FLASH approximates a multi-objective problems as multiple independent problems.  Such an approximations is not appropriate for  models with low constraints and many objectives since solutions
% may be large in number and loosely connected across all objectives (for more technical details on this point, see
% later in this paper).

% \item Two challenge problems for future work.
% \ei
% As stated in the conclusion of this paper, those challenge
% problems are:
% \bi
% \item
% Checking the external validity of our conclusions, that GAs should be deprecated
% for SBSE,  for tasks other than (a),(b),(c),(d);
% \item
% Finding ways to extend FLASH to other kinds of models.
% \ei
% The rest of this paper is structured as follows: we first describe the related work, followed by the description of FLASH. Then, the case-studies used in the paper are described followed by our evaluation.  The paper ends with a discussion on various choices we made while developing FLASH; finally we conclude.  

\section{Related Work}

\subsection{Optimizations from Software Analytics}

One approach towards optimization of SBSE tasks comes from the software analytics literature. 
In this approach,
standard software analytics learners are applied to data
and 
the resulting model is queried in some way to guide the optimizations.
As described below,
examples of this approach include GALE~\cite{krall2015gale}, SWAY~\cite{Nair2016}, and the   CART-based methods~\cite{breiman1984classification} used
by the software product line community~\cite{siegmund2012predicting,nair17}.

Software product line researchers  warn that
developers rarely understand the configurations of their system~\cite{xu2015hey}. 
Hence, they may ignore most of their
configuration options, leaving
 considerable optimization potential
untapped. To address this problem, many researchers
have collected performance data from many different configurations, then applied the CART decision tree learner
to summarize the results into a comprehensible form~\cite{nair2017using, guo2013variability}.
The problem with this approach is the data collection cost-- many configurations must be compiled
and executed to build CART's training set. Accordingly, various research teams have tried
incremental approaches that learn from some stochastically selected sample of the data. For example,
Nair et al.~\cite{nair17} clustered the data then learned from examples selected at random from each cluster. While a useful approach, this method was reliant on CART and hence 
suffer from the limitations of
that algorithm. Specifically: CART learns models for a single goal while the tasks shown in the Section~\ref{sec:intro} require a multiple-goal analysis.

\begin{figure}[!t]
\centering
\small
\hspace{0.4cm}\begin{lstlisting}[mathescape,frame=none,numbers=none]
def SWAY( items, fitness, out) : # "out" is initially the empty list
  if len(items) < enough: 
      out += [items] 
      return out
  east, west = two distant points
  west = fitness(west)
  east = fitness(east)
  goWest = indicatorDominates(east,west)
  goEast = indicatorDominates(west,east)
  if not goWest and not goEast:
       out += [items]
       return out 
  c = dist(east,west)
  for one in items:  # project data onto a line running east, west
      a = dist(x,west)
      b = dist(x,east)   
      one.pos = (a^2 + c^2 - b^2)/(2*c) # cosine rule
  data = sorted(data) # sorted by 'pos'
  mid = len(items)/2  
  # if either dominates, ignore half.  else, recurse on both
  if goEast:
       SWAY(data[:mid], fitness, out)  # cluster east items
  if goWest:
       SWAY(data[mid:], fitness, out) # cluster west items
  return out
\end{lstlisting}
\caption{Framework of SWAY algorithm. For the definition of {\em indicatorDominates}, see Table~\ref{tbl:elite}.
}
\label{fig:sway_frame}  
\end{figure}

Other work from software analytics  handles multi-objective optimization.
Krall et al.~\cite{krall2015gale} approach
optimizations as a  clustering problem. Their GALE algorithm recursively bi-clustered  decisions using an approximation to the first component of principle components analysis.
At each level of the recursion, GALE evaluates  the two most distant decisions, then prunes  half of the data furthest
from a ``better'' decision (and ``better'' is defined using the domination predicates discussed below). For each leaf in the resulting tree, GALE sorts  
 two distant decisions, and mutates
all the examples towards the better end. Subsequent work by Chen and Nair et al.~\cite{Nair2016} showed that GALE's leaf mutation was less important than exploring a  large initial population. Hence, while GALE explored populations of   100 randomly generated decisions, Chen and Nair's SWAY algorithm (shown in Figure~\ref{fig:sway_frame}) used no mutation but just explored  populations  of 10,000 randomly generated decisions. 
The key point of SWAY is when it explores 10,000 decisions, then at every level of its recursion,
it only evaluates two decisions (plus all the decisions found in
the final leaf--
typically $\sqrt{N}$ items). That is, when exploring 10,000 decisions. SWAY terminates after  less than 126 ($2\log_2{10^4} + \sqrt{10^4}$) evaluations.  When compared against evolutionary algorithms,
SWAY proved to be remarkably effective~\cite{Nair2016}. That said, as shown below, the algorithm called FLASH is faster since it uses even fewer evaluations.

\subsection{Bayesian Optimization - ePAL}

Bayesian optimizers (BO)
incrementally update a surrogate model representing a generalization
of all examples seen so far. For example, after processing 20 examples, an BO   generalizes those examples into
some probabilistic regression model
that summarizes   prior beliefs about  the sampled instances. The surrogate model used in practice is usually a Gaussian Process Model~\cite{jamshidi2016uncertainty, zuluaga2016varepsilon} but other approximations have been proposed namely Parzen estimators~\cite{bergstra2011algorithms}, Random Forests~\cite{hutter2011sequential}, just to name a few. 
Once the   surrogate  is available,  
% subsequent fitness function
% ``evaluations'' can be faster (do not query the real world, instead, just query the regression function).
it can be used to decide the next most promising point to evaluate.  This ability to avoid unnecessary model executions (by just exploring a surrogate)
is very useful in the case of models that take minutes to hours to days to execute. 

 \begin{figure}[!t]
\hspace{0.2cm}\begin{lstlisting}[mathescape,numbers=none,frame=none]
  def ePAL(data,  $\varepsilon$, fitness, size0 = 20):  
    some = random.sample(data, size0)
    some = [ fitness(x) for x in some ]
    data = data - some  
    # Till all the points in pop has been either sampled or discarded 
    while len(data) > 0:
      # Build Gaussian Process Model
      model = GP(some)
      # Get uncertainty associated with each point in the pop
      $\mu$, $\sigma$ = model.predict(data)
      # select items to discard based on uncertainty 
      data = data - discard(data,$\mu$, $\sigma$,  $\varepsilon$)
      # Find and evaluate another point.
      some = some + fitness( what2EvaluateNext(data, $\mu$, $\sigma$) )
    return some 
\end{lstlisting}
\caption{\small{The ePAL multi-objective Bayesian optimizer.}}
\label{fig:ePAL}  
\end{figure}
Most prior work on BO   focused on single-goal optimization.
Algorithm~\ref{fig:ePAL} shows 
Zuluaga et al.'s  ePAL~\cite{zuluaga2016varepsilon}, a 
novel multi-objective
extension  of BO.
The algorithm inputs a large number of  unevaluated
data items.
Before entering  its main loop, ePAL applies the fitness
function to an initial small sample of 20 data items.
Then, until it exhausts all the data, the evaluated data
are used to build a Gaussian process model. The model
is then applied to the remaining data to learn  $\mu,\sigma$
of the model predictions. ePAL then removes all  $\varepsilon$-dominated points:
$y$ is discarded due to $x$ if $\mu_x + \sigma_x$  $\varepsilon$-dominates $\mu_y - \sigma_y$, where  
$x$  $\varepsilon$-dominates $y$ if $x + \varepsilon \succeq  y$
and  ``$\succeq$'' is   binary domination- see Equation~\ref{eq:bdom} in Table~\ref{tbl:elit}.
ePAL also uses its model to select the next most informative data item
to evaluate. For this task, it selects the instance which is furthest away
from the known $\mu+\sigma$ of any objective.

While an interesting technology in some test domains,  Bayesian optimization has certain limitation.
Building Gaussian process models can be very challenging for for high dimensional data. So far,
the state-of-the-art in this arena is the optimization
of models with around ten decisions ~\cite{wang2016bayesian}.

\subsection{Evolutionary Algorithms}\label{sec:ga}
EAs (also referred to as MOEA-Multi-Objective EA) evolve a population of solutions, guided by a fitness function, as follows.
{\em Step 1:}  Generate an initial population of solutions using an initialization policy for example random sampling.
{\em Step 2:}  Evaluate each solution using a problem specific fitness function.
{\em Step 3:}  Repeat the following.
    \begin{itemize}
        \item Create a new population using some problem specific reproduction operators; e.g. new individuals are formed by cross-over parts of pairs of parents.
        \item Evaluate the new  population via the fitness function.
        \item Select solutions from a new population for the next generation. This selection is done using an elitist strategy which mimics the ``survival of the fittest''. Table~\ref{tbl:elite} discusses different kinds of elitism.
    \end{itemize}
% \textcolor{red}{
% (Aside: Note that SWAY is not a GA-- it performs {\em Step1} and 
% {\em Step2} then, only once, its
% runs that recursive bi-clustering process to select some individuals. Then
% SWAY terminates without looping through {\em Step3}.)} -- is this required?

{\em Domination counters}    compute $\Omega_d(x)$ which is a sort order for  individuals in a population  based on how many other individuals $y$ that are worse than $x$, according to a domination definition $d$. For a sample of different domination definitions, see   Table~\ref{tbl:elit}. 
One nuance of multi-objective optimization is that when a population  is sorted by dominance, there may be more than one ``best'' individual. 
This is especially true when using  the {\em binary domination} $\Omega_B$ definition of
Equation~\ref{eq:bdom} in Table~\ref{tbl:elite}. Hence, 
algorithms like NSGA-II~\cite{deb2002fast}, rely on a fast ``non-dominating sorting'' procedure to divide the population into ``bands'' with the property that items in the first band are better than those in the second, etc. Accordingly, when we seek several useful individuals, we will use   $\Omega_B$.
However, when we want
a single ``best'' individual, we will use another dominance counter which is known to return a single individual as ``best''
(the  $\Omega_I$ counter based on indicator
dominance-- see  Equation~\ref{eq:cdom}
of Table~\ref{tbl:elite}).

\begin{table}
\caption{Three kinds of EA elitist strategies.}\label{tbl:elit}
{\small
\begin{tabular}{|p{.95\linewidth}|}\hline
{\em 1. Decomposition-based algorithms}
such as MOGLS~\cite{ishibuchi1998multi} and  MOEA/D~\cite{zhang2007moea}
divide the problem into a set of sub-problems, which are solved simultaneously in a collaborative manner.  For example, at start-up,
MOEA/D generates over-lapping clusters of the  population to find individuals with similar goal. If any member of a cluster finds a better solution, then it broadcasts the direction of that improvement to all individuals of all its
containing clusters.\\\hline

{\em 2. Pareto dominance-based algorithms} such as
 NSGA-II~\cite{deb2002fast}, PAES~\cite{knowles2000approximating} and SPEA2~\cite{zitzler2001spea2}
 use  {\em  binary domination} to select solutions for the successive generations.
When exploring a complex set of competing goals, there may be no best solution that is best over all objectives. Hence,
to declare that one solution is ``better'' than another,
all objectives must be polled separately. Given two vectors of decisions $x,y$ with associated objectives $o_x, o_y$,
then $x$ is binary dominant over $y$ when:
\begin{equation}\label{eq:bdom}
    \begin{split}
             \forall o_j  \in \textit{objectives}\;\mid\;
             \neg ( o_{j,x} \prec o_{j,y}) \mid \\
         \wedge \exists o_j \in \textit{objectives} \;\mid\; o_{j,x} \succ o_{j,y}\mid 
    \end{split}
\end{equation}
where \textit{obj} are the objectives and ($\succeq,\succ$) tests if an objective score in one individual is (no worse, better) than  the other.
Pareto dominance-based algorithm  are used in tandem with niching operators to preserve the diversity among the solutions. For example, if binary domination selects
 too many candidates for the next generation, NSGA-II employs a crowd-pruning heuristic to stop solutions clumping together to closely.\\\hline

{\em 3. Indicator-based algorithms} such as 
HypE~\cite{bader2011hype} and IBEA~\cite{zitzler2004indicator}
 work by establishing a complete order among the solutions using a single scalar metric $M$ like hypervolume etc. 
 For example, in IBEA, $x$ indicator dominates over $y$ if
 
\begin{equation}\label{eq:cdom}
    \begin{array}{rcl}
    x \succ y & =& \textit{M}(y,x) > \textit{M}(x,y)\\
    \textit{M}(x,y)& = &\sum_j^n -e^{\Delta(j,x,y,n)}/n\\
    \Delta(j,x,y,n) & = & w_j(o_{j,x}  - o_{j,y})/n
    \end{array}
\end{equation}
In the above $w\in \{-1,1\}$  and represents the $n$ objectives that need to be minimized or maximized. \\\hline
\end{tabular}}
\label{tbl:elite}
\end{table}

Two problems with EAs are their {\em long runtimes} and {\em comprehensibility}.
In practice, EA evaluates far more individuals than
other algorithms such as  SWAY, ePAL, or FLASH algorithm discussed in the next section. For
example,
when completing tasks (a), (b), (c), (d) from our introduction,
(ePAL, EAs) require (tens, thousands) of evaluations, respectively.
For another (very extreme) example,
Wang et al.~\cite{wang2013searching} needed 15 years of CPU time for their EAs to tune the parameters of their software clone detection tools.
More typically, Harman~\cite{harmanP} comments
on the problems of evolving a test suite for software
if every candidate solution requires a time consuming
execution of the entire system: such test suite generation
can take weeks of execution time. 

As to the {\em comprehension} problem, when EAs terminate, they return all the individuals in the final population. Valerdi notes that,
without automated tools, it can take days for human experts to review just a few dozen examples. In that same time, an automatic tool can explore thousands to billions more solutions. Humans can find it an overwhelming task just to certify the correctness of conclusions generated from so many results. Verrappa and Leiter warn that:
\begin{quote}
{\em ... for industrial problems, these algorithms generate (many) solutions which make the task of understanding them and selecting one amongst them difficult and time consuming.}~\cite{veerappa2011understanding}
\end{quote}

% fitness function can be called fewer times. Consider an optimizer trying to learn a good configuration of the Makefile parameters within MySQL. As it experiments with different configurations, it must evaluate each one. If that fitness function is ``reduce the runtime of the MySQL test suite'', then one
% evaluation of that function requires:
% \bi
% \item
% A full recompilation of MySQL;
% \item
% Followed by a full execution of its test suite.
% \ei
% Now consider the BPO approach. After evaluating few different configurations (20 is a typical number~\cite{zuluaga2013active}),
% the BPO generalizes that experience into probabilistic regression model. Once that generalization is available, then future ``evaluations'' could be either {\em actual}, or {\em mocked} where the latter means use the probabilistic regression model to predict what might happen if we actually recompiled MySQL and ran the test suite.  This approach is useful
% since it can be orders of magnitude slower to conduct 
% and actual evaluation than to mock
%the evaluation using the probabilistic regression model.

\subsection{FLASH: A Hybrid Algorithm}
Our understanding of the literature is that the optimization work 
in software analytics, machine learning, and evolutionary algorithms has evolved
on mostly separate lines. But, clearly, these methods were all evolved
to achieve similar goals. Therefore, it is reasonable to ask if there is any advantage in combining ideas from these different research directions.

Considerations such as these lead to the design of FLASH as a combination of other methods:
\bi
\item From the EA community, we took the \textit{dominance counters} $\Omega_B$ and
$\Omega_I$ since these can find us find good candidate(s) within a large space of multiple objective options.
\item
From the Bayesian optimization work, we took the technique of \textit{minimizing the calls to the fitness functions}; i.e. build a model from the current evaluations then use that model as a surrogate for the real world evaluations. 
\item
From the software analytics community, we took \textit{decision tree learning with CART}
since such trees offer a succinct representation of multiple examples.
Another advantage of CART is that this algorithm does scale to large dimensional models (whereas the Gaussian process models used in, say, ePAL struggle to build models for more than ten dimensions).
\ei
Standard CART has the disadvantage that it only builds models for a single goal.
Hence,  FLASH uses CART to build a separate tree for each objective in a multi-objective problem.

The resulting algorithm is shown in  Figure~\ref{fig:flash_frame}. 
For each loop of the algorithm, the
{\em best} set of solutions is repeatedly
pruned by  $\Omega_B$ (the NSGA-II fast non-dominating sort algorithm). FLASH
tries to grow {\em best}  by running its CART
models (one for each objective) over all the data in order to find the single
example that looks most promising. However, if that new example fails to grow the {\em best} set, FLASH looses a life.

FLASH only executes a full evaluation, once per cycle of its {\bf while} loop; i.e. measured in terms of number of evaluations,
FLASH should run very fast. Further, since it uses CART and not Gaussian process models, it should scale to models with much more than ten dimensions. 

FLASH offers certain novel innovations over other
work that tried optimizing BO by replacing 
Gaussian process models with other learners. 
Researchers exploring methods to optimize BO beyond ten decisions typically assumed single
objective tasks~\cite{bergstra2011algorithms, hutter2011sequential, thornton2013auto}.  
Also, for FLASH, we strive to generate succinct descriptions of its processing (one small decision tree per objective).
Other researchers never even attempt to explore
comprehensible so for their BO variants, they use random forests-- which typically
generate 10s to 100s of trees~\cite{hutter2011sequential}.

 \begin{figure}
\small
\hspace{0.2cm}\begin{lstlisting}[mathescape,frame=none,numbers=none]
def FLASH(data, fitness, size0=20, lives=10):
    best = random.sample(data, size0)
    best = [ fitness(x) for x in some ]
    while lives > 0:
        for o in objectives: # build one CART model per objective
            model[o] = CART(some, target=o)
        more =  best + fitness( what2EvaluateNext(data, some) ),
        tmp = $\Omega_B$(more)  # non-dominated sort
        if tmp == best: # does the new item grow the set of "best" ideas?
            lives = lives - 1 # if no, then lose a life
        else:
            best = tmp # else, we have a new set of "best ideas"
   return best

def what2EvaluateNext(items, some):    
    for one in items:
       for o in objectives: # score each item on each CART model
         one[o] = model[o].predict(one)
    return $\Omega_I$(  # select the best one, not in "some" within the
             $\Omega_B$(items - some)) # good individuals found by ``$\Omega_B$''
\end{lstlisting}
\caption{Framework of the FLASH algorithm. $\Omega_B$ is the NSGA-II
fast non-dominating sort procedure that uses binary domination to return a set
of useful individuals. $\Omega_I$ 
uses indicator dominance to return a single best individual.}
\label{fig:flash_frame}  
\end{figure}

\begin{figure*}[!t]
\centering
\scriptsize
    \setlength{\tabcolsep}{3.5pt} % Default value: 6pt
    \renewcommand{\arraystretch}{1} % Default value: 1
\resizebox{1\linewidth}{!}{\begin{tabular}{|p{1.2cm}|p{1.35cm}|p{0.5cm}|c|c|m{8.5cm}|p{7cm}|p{2.5cm}|p{0.5cm}|}
% \begin{tabular}{|l|l|l|l|l|l|l|l|l|}
\hline
\textbf{Family}                                                                                          & \textbf{Variants}     & \textbf{Abbr} & \textbf{\#Dec.}          & \textbf{\#Obj.}        & \textbf{Description}                                                                                                                                                                                  & \textbf{Decisions}                                                                                                                                                                  & \textbf{Objectives}                                                                & \textbf{\begin{tabular}[c]{@{}l@{}}Prev\\ Used\end{tabular}} \\ \hline
\multirow{15}{*}{\begin{tabular}[c]{@{}l@{}}Configuration\\ Control\\ \end{tabular}}               & wc-c1-3d-c1  & SS1  & 3                   & \multirow{15}{*}{2} & Word Count is executed by varying 3 configurations of Apache Storm on cluster C1                                                                                                             & max\_spout,  spliters, counters                                                                                                                                            & Throughput \& Latency                                                     & \cite{jamshidi2016uncertainty}                          \\ \cline{2-4} \cline{6-9} 
                                                                                                & sort-256     & SS2  & 3                   &                     & The design space consists of 206 different hardware implementations of a  sorting network for 256 inputs                                                                                     & Not specified                                                                                                                                                              & Area \&Throughput                                                         & \cite{zuluaga2016varepsilon}                            \\ \cline{2-4} \cline{6-9} 
                                                                                                & wc-c3-3d-c1  & SS3  & 3                   &                     & Word Count is executed by varying 3 configurations of Apache Storm on cluster C3                                                                                                             & max\_spout, spliters, counters                                                                                                                                             & Throughput \& Latency                                                     & \cite{jamshidi2016uncertainty}                          \\ \cline{2-4} \cline{6-9} 
                                                                                                & wc+wc-3d-c4  & SS4  & 3                   &                     & Word Count is executed, collocated with Word Count task, by varying 3 configurations of Apache Storm on cluster C3                                                                           & max\_spout, spliters, counters                                                                                                                                             & Throughput \& Latency                                                     & \cite{jamshidi2016uncertainty}                          \\ \cline{2-4} \cline{6-9} 
                                                                                                & wc-3d-c4     & SS5  & 3                   &                     & Word Count is executed by varying 3  configurations of Apache Storm on cluster C4                                                                                                            & max\_spout, spliters, counters                                                                                                                                             & Throughput \& Latency                                                     & \cite{jamshidi2016uncertainty}                          \\ \cline{2-4} \cline{6-9} 
                                                                                                & wc+rs-3d-c4  & SS6  & 3                   &                     & Word Count is executed, collocated with Rolling Sort task, by varying 3 configurations of Apache Storm on cluster C3                                                                         & max\_spout, spliters, counters                                                                                                                                             & Throughput \& Latency                                                     & \cite{jamshidi2016uncertainty}                          \\ \cline{2-4} \cline{6-9} 
                                                                                                & wc+sol-3d-c4 & SS7  & 3                   &                     & Word Count is executed, collocated with SOL task, by varying 3 configurations of Apache Storm on cluster C3                                       & max\_spout, spliters, counters                                                                                                                                             & Throughput \& Latency                                                     & \cite{jamshidi2016uncertainty}                          \\ \cline{2-4} \cline{6-9} 
                                                                                                & noc-CM-log   & SS8  & 4                   &                     & The design space consists of 259 different implementations of a tree-based network-on-chip,  targeting application specific circuits (ASICs) and multi-processor system-on-chip designs      & Not specified                                                                                                                                                              & Energy \& runtime                                                         & \cite{zuluaga2016varepsilon}                            \\ \cline{2-4} \cline{6-9} 
                                                                                                & wc-5d-c5     & SS9  & 5                   &                     & Word Count is executed by varying 5 configurations of Apache Storm on cluster C3                                                                                                             & spouts, splitters, counters, buffer-size, heap,                                                                                                                            & Throughput \& Latency                                                     & \cite{jamshidi2016uncertainty}                          \\ \cline{2-4} \cline{6-9} 
                                                                                                & rs-6d-c3     & SS10 & 6                   &                     & Rolling Sort is executed by varying 6 configurations of Apache Storm on cluster C3                                                                                                           & spouts, max\_spout, sorters, emitfreq, chunksize, message\_size                                                                                                            & Throughput \& Latency                                                     & \cite{jamshidi2016uncertainty}                          \\ \cline{2-4} \cline{6-9} 
                                                                                                & wc-6d-c1     & SS11 & 6                   &                     & Word Count is executed by varying 6 configurations of Apache Storm on cluster C1                                                                                                             & spouts, max\_spout, sorters,  emitfreq, chunksize,  message\_size                                                                                                          & Throughput \& Latency                                                     & \cite{jamshidi2016uncertainty}                          \\ \cline{2-4} \cline{6-9} 
                                                                                                & llvm         & SS12 & 11                  &                     & The design space consists of 1023 different compiler settings for the LLVM compiler framework. Each setting is specified by d = 11 binary flags.                                             & time\_passes, gvn, instcombine, inline, ...,  ipsccp, iv\_users, licm                                                  & Performance \& memory footprint & \cite{zuluaga2016varepsilon}                            \\ \cline{2-4} \cline{6-9} 
                                                                                                & Trimesh      & SS13 & 13                  &                     & Configuration space of Trimesh, a library to manipulate triangle meshes                                                                                                                      & F, smoother, colorGS, relaxParameter, V, Jacobi, line, zebraLine, cycle, alpha, beta, preSmoothing, postSmoothing                                                          & Number Iterations \& TimeToSolution                                        & \cite{siegmund2012predicting}                           \\ \cline{2-4} \cline{6-9} 
                                                                                                & X264-DB      & SS14 & 17                  &                     & Configuration space of X-264 a video encoder                                                                                                                                                 & no\_mbtree, no\_asm, no\_cabac, no\_scenecut, ..., keyint, crf, scenecut, seek, ipratio                 & PSNR \& Energy                                                            & \cite{siegmund2012predicting}                           \\ \cline{2-4} \cline{6-9} 
                                                                                                & SaC          & SS15 & 59                  &                     & Configuration space of SaC                                                                                                                                                                   & extrema, enabledOptimizations, disabledOptimizations, ls, dcr, cf, lir, inl, lur, wlur, ... maxae, initmheap, initwheap                                                    & compile-exit, compile-real                                                & \cite{siegmund2012predicting}                           \\ \hline
\multirow{4}{*}{\begin{tabular}[c]{@{}l@{}}Release\\  Planning\\  (MONRP)\\ \end{tabular}}         & 50-4-5-0-110 & N1   & \multirow{4}{*}{50} & \multirow{4}{*}{3}  & The project has 50 requirements with 4 releases. It has 5 clients and requirement has no dependency on each other and is over-funded (10\%).                   & \multirow{4}{*}{\shortstack[l]{Each decision represents the attributes of the  requirements such as \\ risk, cost}} & \multirow{4}{*}{Risk, Satisfaction \& Cost}                               & \cite{chen2017beyond}                                   \\ \cline{2-3} \cline{6-6} \cline{9-9} 
                                                                                                & 50-4-5-0-90  & N2   &                     &                     & The project has 50 requirements with 4 releases. It has 5 clients and requirement has no dependency on each other and is 90\% underfunded.                         &                                                                                                                                                                            &                                                                           & \cite{chen2017beyond}                                   \\ \cline{2-3} \cline{6-6} \cline{9-9} 
                                                                                                & 50-4-5-4-90  & N3   &                     &                     & The project has 50 requirements with 4 releases. It has 5 clients and requirement has 4\% of the requirements are dependent on each other and is 90\% underfunded &                                                                                                                                                                            &                                                                           & \cite{chen2017beyond}                                   \\ \cline{2-3} \cline{6-6} \cline{9-9} 
                                                                                                & 50-4-5-4-110 & N4   &                     &                     & The project has 50 requirements with 4 releases. It has 5 clients and requirement has no dependency on each other and is overfunded (10\%).                  &                                                                                                                                                                            &                                                                           & \cite{chen2017beyond}                                   \\ \hline
\end{tabular}}
    \caption{SS* and MONRP are {\em constrained models.}}
    \label{fig:constrained_case_studies}
\end{figure*}

\begin{figure*}[!t]
\centering
\scriptsize
    \setlength{\tabcolsep}{3.5pt} % Default value: 6pt
    \renewcommand{\arraystretch}{1} % Default value: 1
\resizebox{1\linewidth}{!}{\begin{tabular}{|p{1.2cm}|p{1.35cm}|p{0.5cm}|c|c|m{8.5cm}|p{7cm}|p{2.5cm}|p{0.5cm}|}
% \begin{tabular}{|l|l|l|l|l|l|l|l|l|}
\hline
\textbf{Family}                                                                                          & \textbf{Variants}     & \textbf{Abbr} & \textbf{\#Dec.}          & \textbf{\#Obj.}        & \textbf{Description}                                                                                                                                                                                  & \textbf{Decisions}                                                                                                                                                                  & \textbf{Objectives}                                                                & \textbf{\begin{tabular}[c]{@{}l@{}}Prev\\ Used\end{tabular} }\\ \hline

\multirow{4}{*}{\begin{tabular}[c]{@{}l@{}}Model of\\ Agile \\ Development\\  (POM) \end{tabular}} & A            & P1   & \multirow{4}{*}{9}  & \multirow{4}{*}{3}  & The simulated project has a fairly large team, where only a 2-10\% of the teams are affected by criticality.                                                                                 & \multirow{4}{*}{\shortstack[l]{Culture, Criticality, Criticality-Modifier, Initial known,\\ Inter-dependency,  Dynamism, Size, Plan, Team Size}}  & \multirow{4}{*}{\shortstack[l]{Completion Rates, Idle\\ Cost  \& Overall Cost }}            & \cite{krall2015gale}                                    \\ \cline{2-3} \cline{6-6} \cline{9-9} 
                                                                                                & B            & P2   &                     &                     & The simulated project has a small team, where most of teams, 80-90\% of teams are affected by criticality.                                                                                   &                                                                                                                                                                            &                                                                           & \cite{krall2015gale}                                    \\ \cline{2-3} \cline{6-6} \cline{9-9} 
                                                                                                & C            & P3   &                     &                     & The simulated project has a fairly large team. The project is very dynamic and new requirements are added to the project frequently.                                                         &                                                                                                                                                                            &                                                                           & \cite{krall2015gale}                                    \\ \cline{2-3} \cline{6-6} \cline{9-9} 
                                                                                                & D            & P4   &                     &                     & \begin{tabular}[c]{@{}l@{}}The simulated project has a small team. In this project, the teams are highly\\ dependent on each other.\end{tabular}                                            &                                                                                                                                                                            &                                                                           & \cite{krall2015gale}                                    \\ \hline
\multirow{5}{*}{\begin{tabular}[c]{@{}l@{}}Process \\ Modelling \\(XOMO) \\  \end{tabular}}                   & Ground       & X1   & \multirow{5}{*}{27} & \multirow{5}{*}{4}  & Simulate the JPL Ground Software                                                                                                                                                             & \multirow{5}{*}{\shortstack[l]{aa, sced, cplx, site, resl, acap, etat, rely,,Data, prec, pmat, aexp,\\ flex, pcon, tool, time,stor, docu, b, plex, pcap, kloc,\\ ltex, pr, ruse, team, pvol}} & \multirow{5}{*}{\shortstack[l]{ Effort, Months, \\Defects, Risks}}                           & \cite{krall2015gale}                                    \\ \cline{2-3} \cline{6-6} \cline{9-9} 
                                                                                                & OSP          & X2   &                     &                     & Simulates the orbital space plane guidance navigation and control                                                                                                                            &                                                                                                                                                                            &                                                                           & \cite{krall2015gale}                                    \\ \cline{2-3} \cline{6-6} \cline{9-9} 
                                                                                                & O2           & X3   &                     &                     & Simulates the orbital space plane guidance navigation and control (version 2)                                                                                                                &                                                                                                                                                                            &                                                                           & \cite{krall2015gale}                                    \\ \cline{2-3} \cline{6-6} \cline{9-9} 
                                                                                                & ALL          & X4   &                     &                     &   Simulates   Flight Software  from the Jet Propulsion Lab                                                                                                                                                                                           &                                                                                                                                                                            &                                                                           & \cite{krall2015gale}                                    \\ \cline{2-3} \cline{6-6} \cline{9-9} 
                                                                                                & Flight       & X5   &                     &                     & Simulates   Flight Software  from the Jet Propulsion Lab                                                                                                                                                          &                                                                                                                                                                            &                                                                           & \cite{krall2015gale}                                    \\ \hline
\end{tabular}}
    \caption{  POM and XOMO are {\em unconstrained models. }}
    \label{fig:unconstrained_case_studies}
\end{figure*}

\section{Experimental Materials and Methods}

The rest of this paper offers experiments to comparatively evaluate FLASH versus other algorithms using the tasks described in our introduction.

\subsection{Experimental Design Principle: ``Precedented  plus Two''}
Our reading of the optimization literature is that
there exists a very large number
of optimizers, models, evaluation criteria
that could be used to design experiments with optimizers. No single paper can explore all combinations of the above.
Hence, we need some guiding principle for experimental design. 

The principle used here is as follows: {\em
precedented plus two}. 
As to ``{\em plus two}'', we think important to not just use past work, but also to extend some parts of that work. Accordingly, to
our design, we add one new treatment we wish to test (in this case, FLASH)
plus a second new treatment that was rarely used before. 
In this case, we will  apply the  MOEA/D evolutionary algorithm (described in Table~\ref{tbl:elite}) to many of our models. MOEA/D was chosen since it is an example of the newer generation of EA algorithms.

As to ``{\em precendented}'', this means  our experimental methods need to be justified via some prior precedent in the literature.
For example, to select our statistical hypothesis test
methods, we use  the Scott-Knott  effect-size  endorsed
by Mittas and Angelis at TSE'13~\cite{mittas13}
and by Hassan et. al. at ICSE'15~\cite{7194626}.  Within Scott-Knot,
we use the A12 non-parametric  effect size test (to detect, then reject,
trivially small differences) since A12 was endorsed
by Acura \& Briand at ICSE'11 in their paper {\em Statistical tests to assess
randomized algorithms in SE}~\cite{arcuri2011practical}.

Another paper used to design our experiments is 
{\em Practical guide to select quality indicators for   search algorithms } recently published in   ICSE'16~\cite{wang2016practical}.
Following their advice, we use two measures to assess the success of a multi-objective optimizer:
\bi
\item
The Generational Distance (GD)~\cite{van1999multiobjective} measures the closeness  of the solutions from by the optimizers to the 
{\em Pareto frontier} i.e. the  actual set of non-dominated solutions. 
\item
The Inverted Generational Distance (IGD)~\cite{coello2004study}
is the mean distance from points on the  Pareto frontier to its nearest  point in the set returned by the optimizer.
\ei
Note that, for both measures, {\em smaller} values are {\em better}.
Also,
according to
Coello et al.~\cite{coello2004study}, IGD is a better measure
of how well an optimizer's solutions {\em spreads}   across the space of all known solutions.
Further, for  models other than SS*, obtaining the PF is infeasible in practice~\cite{deb2005scalable}. Thus, to obtain a  Pareto frontier, we apply $\Omega_B$
to all solutions found by any algorithms for one model.

Moving on the selection of models,
applying ``{\em precendented}'' principle, we use the
POM, XOMO, MONRP, SS* models, respectively, to
address the {\em Introduction}'s tasks (a), (b), (c), (d) since:
\bi
\item
POM and XOMO were used extensively
by  Chen et al. in their work on SWAY~\cite{Nair2016, krall2015gale}.  
\item
  MONRP is widely used in the next-release planning community~\cite{chen2017beyond}.
\item
The SS* models were used by Zuluga et al. in their publications that report results for BO~\cite{zuluaga2016varepsilon, jamshidi2016uncertainty}.
\ei
For  details on these models, see 
Figure~\ref{fig:constrained_case_studies} and
Figure~\ref{fig:unconstrained_case_studies} and  \S\ref{sec:models}
(below).

Once the test models are known, the next step is to select optimizers that have been previously applied to those models. For example, Chen et al. made extensive use of NSGA-II and SPEA2 for their studies of POM and XOMO.
Also,
ePAL was applied by Zuluga et al. to the SS* models so we will do the same. Note that we we will run two version for ePAL:
\bi
\item $\mathit{ePAL}(\varepsilon=0.01)$ and
\item $\mathit{ePAL}(\varepsilon=0.3)$
\ei
These ePAL versions  represent two extremes of ePAL from most
cautious ($\varepsilon=0.01$) to most careless ($\varepsilon=0.3$).

Other aspects of our  design were designed in response with the specific features of our experiments. For example,
all our algorithms were written in Python except for ePAL, which uses the Matlab code from the  Zuluga et al. group.
Since we are comparing implementations in different languages,
we do not measure ``speed'' in terms of runtimes. Rather, we use ``number of evaluations''
to measure speed since that is a language-independent feature.

 Another specific aspect of our design is how we measure model {\em comprehensibility}. 
We assume that software engineers will want to browse, understand and audit the results of any
 algorithm that has the presumption to tell them to change this or that.
 One way to present the results of an optimizer is to generate a {\em domination tree}, as follows:
 \bi
 \item
 Take all the examples
 ever evaluated by an optimizer; 
 \item
 Score each individual  by $|\Omega_I(x)|$, which is the number
 of other individuals dominated by $x$; 
 \item
 Use the CART decision tree learner~\cite{breiman1984classification} to summarize the
 decisions that lead to difference domination scores.
 \ei
  In
 this paper, we will say {\em domination trees} with {\em fewest} nodes and leaves
 are {\em more} comprehensible. 
 For example, Figures~\ref{fig:dom1} and ~\ref{fig:dom2} show two trees generated from
 the examples evaluated by FLASH and ePAL while optimizing the LLVM (SS12)  model of 
 Figure~\ref{fig:unconstrained_case_studies}.
 Note that one of these trees is far smaller  and hence easier
 to browse, understand, and audit.

 \begin{figure}[!b]
\hspace{0.2cm}\begin{lstlisting}[
escapechar=@,
basicstyle=\scriptsize\ttfamily,
keepspaces=true,
linewidth=3.25in,
frame=none,
numbers=right]
@\bh@sccp=0@\eh@
|    print_used_types=0
|    |    ipsccp=0 (6.5)
|    |    ipsccp=1
|    |    |    x[10]=0 (12) 
|    |    |    x[10]=1 (19) 
|    @\bh@print_used_types=1@\eh@
|    |    ipsccp=0 (14)
|    |    @\bh@ipsccp=1@\eh@
|    |    |    time_passes=0  (24)
|    |    |    @\bh@time_passes=1@\eh@
|    |    |    |    jump_threading=0 (28)
|    |    |    |    @\bh@jump_threading=1@\eh@ (31)
sccp=1
|    ipsccp=0 (1.5) 
|    ipsccp=1 (7.5) 
\end{lstlisting}
\caption{\small{A domination 
tree. Optimizer = FLASH; model= LLVM, from Figure~\ref{fig:constrained_case_studies}.
Numbers in brackets (e.g. ``(31)'') show how many other individuals are dominated by individuals in a particular leaf.
For example, on the last line, there is a leaf whose individuals dominate (on average)
31 others.
The branch marked in gray leads to the branch with greatest score.
}}\label{fig:dom1}  
\end{figure}

\subsection{Model Details}\label{sec:models}

Two aspects of note about the models used in this study
are the  {\em number of decisions}
and the {\em presence of constraints} on valid decisions:
\bi
\item
Figure~\ref{fig:constrained_case_studies} shows our models with
constraints;
\item
Figure~\ref{fig:unconstrained_case_studies} shows our models without constraints;
\item The {\em \#Decs} column of these two figures shows how many 
decisions are used by these models.
\ei
These aspects are important since, as seen below, these
aspects  determine which of our six optimizers
(ePAL, FLASH, SWAY, NSGA-II, SPEA2, MOEA/D)
 succeed on different models.

Our first set models are the ``SS*'' models of Figure~\ref{fig:constrained_case_studies}. These models are
related to 
configuration control for software system. Adjusting the choice
points (configuration) in these models alters the performance objectives (listed on the right-hand-side -- Objectives, of Figure~\ref{fig:constrained_case_studies}).
These SS* models come from (i) Zuluaga et.al~\cite{zuluaga2016varepsilon} (SS2, SS8, SS12) and (ii) Jamshidi et al.~\cite{jamshidi2016uncertainty} (all the rest). Zuluaga et al. explored two chip-design problems (SS2, SS8) and a software configuration problem (SS12), where evaluation of a single point is very expensive. Jamshidi et al.~\cite{jamshidi2016uncertainty} ran three benchmarks (Word Count, Rolling Sort and SOL) on Apache Storm on clusters with different cluster configuration).

 \begin{figure}[!t]
\hspace{0.2cm}\begin{lstlisting}[
escapechar=@,
basicstyle=\scriptsize\ttfamily,
keepspaces=true,
linewidth=3.25in,
frame=none,
numbers=right]
ipsccp=0
|    iv_users=0
|    |    sccp=0
|    |    |    print_used_types=0
|    |    |    |    jump_threading=0
|    |    |    |    |    time_passes=0
|    |    |    |    |    |    instcombine=0 (16)
|    |    |    |    |    |    instcombine=1 (12)
|    |    |    |    |    time_passes=1 (10)
|    |    |    |    jump_threading=1
|    |    |    |    |    simplifycfg=0 (32)
|    |    |    |    |    simplifycfg=1 (33)
|    |    |    print_used_types=1
|    |    |    |    |inline=0 (47)
|    |    |    |    |inline=1 (43)
|    |    sccp=1
|    |    |    print_used_types=0 (1)
|    |    |    |print_used_types=1 (7)
|    iv_users=1
|    |    sccp=0
|    |    |    print_used_types=0
|    |    |    |    jump_threading=0 (30)
|    |    |    |    jump_threading=1 (42)
|    |    |    print_used_types=1
|    |    |    |    inline=0 (56)
|    |    |    |    inline=1 (62)
|    |    sccp=1
|    |    |    instcombine=0 (26)
|    |    |    instcombine=1 (33)
@\bh@ipsccp=1@\eh@
|    @\bh@sccp=0@\eh@
|    |    print_used_types=0
|    |    |    iv_users=0
|    |    |    |    jump_threading=0 (50)
|    |    |    |    jump_threading=1 
|    |    |    |    |    instcombine=0 (53)
|    |    |    |    |    instcombine=1 (54)
|    |    |    iv_users=1
|    |    |    |    simplifycfg=0 (59.5)
|    |    |    |    simplifycfg=1 
|    |    |    |    |    time_passes=0 (63)
|    |    |    |    |    time_passes=1 (66)
|    |    @\bh@print_used_types=1@\eh@
|    |    |    iv_users=0
|    |    |    |    time_passes=0 (69)
|    |    |    |    time_passes=1
|    |    |    |    |    instcombine=0 (73)
|    |    |    |    |    instcombine=1 (71)
|    |    |    @\bh@iv_users=1@\eh@
|    |    |    |    gvn=0 (76)
|    |    |    |     @\bh@gvn=1 (79)@\eh@
|    sccp=1
|    |    print_used_types=0
|    |    |    jump_threading=0 (3)
|    |    |    jump_threading=1 (16)
|    |    print_used_types=1
|    |    |    iv_users=0 (35)
|    |    |    iv_users=1 (50)
\end{lstlisting}
\caption{\small{Another domination tree. Same model (LLVM) and
format as Figure~\ref{fig:dom1} but the optimizer is ePAL.
This tree is larger than Figure~\ref{fig:dom1}
(and hence harder to understand) since ePAL evaluated more examples than FLASH.
}}\label{fig:dom2}  
\end{figure}

The other constrained models are   Multi-Objective Next Release Planning problem  (MONRP). These are concerned with defining which requirements should be implemented for the next release of a software. The flavor of MONRP used in this paper considers (maximizing) combination of importance of features and corresponding risk, (minimizing) cost and (maximizing) satisfaction. Thus the multi-objective Next Release Problem can be formalized as  the process of maximizing $f_1,f_2$ and minimizing $f_3$,defined as follows:

{\scriptsize \[\begin{array}{c}
  f_1=\sum\limits_{i=1}^N (\mathit{score_i} . (P - x_i + 1) - \mathit{risk}_i.x_i).y_i \\
     \mathit{and}\; f_2=\sum\limits_{i=1}^N r_i \;  \mathit{and}\;
     f_3=\sum\limits_{i=1}^N c_i.y_i 
\end{array}\]}
subject to:

{\scriptsize \begin{center}
   $ \sum\limits_{i=1}^N \mathit{cost}_i. f_{i,k} \le \mathit{budgetRelease}_k, \forall k \in {1, ..., P}$\\
   $x_b \le x_a, \forall(r_a \rightarrow r_b), (\mathit{where}\; r_a, r_b \to R)$
\end{center}}
where, 
    $score_i = \sum\limits_{j=1}^M wt_j. (\mathit{importance(c_j, r_i)}$, the Boolean variable $y_i$ indicates whether requirement $r_i$ will be implemented in some release, variable $x_i$ indicates the number of the release where requirement $r_i$ is to be implemented. $\mathit{budgetRelease}_k$ refers to the available budget and $P$ represents the number of releases. We explore four variants of MONRP, ranging from the least constraint to the most constraint. For example, problem variant MONRP-50-4-5-4-110 refers to a scenario where a software project has 50 requirements and 4 releases. The project involves development of a software for 5 clients, delivered in the form of 4 releases using 110\% of the actual cost. For more details refer to~\cite{chen2017beyond}.

Our next two models are unconstrained.
POM and XOMO, were designed for software process
control for agile and waterfall systems, respectively.
POM was based on the work of
Turner and Boehm~\cite{boehm2003balancing} who  observed that agile managers struggle to balance idle rates of developers, completion rates and over all cost of the project. POM is a model to compute completion rates, idle times and overall costs. POM models the agile process by considering a set of inter-dependent requirements. Each requirement consists of a priority value and corresponding cost along with list of dependent requirements, which need to be satisfied before completing the requirements. Since POM models an agile environment, the cost and value of the requirements are constantly changing until the completion of the requirement. The POM model considers minimizing the man hours spent on developing a requirement, salary of the developers and the idle times. We explore four variants of POM ranging for a small highly critical project to large project which is very dynamic in nature. For more details refer to~\cite{port08,chen2017beyond}.

\begin{figure*}[!th]
\centering
    \includegraphics[width=\linewidth, height=4cm]{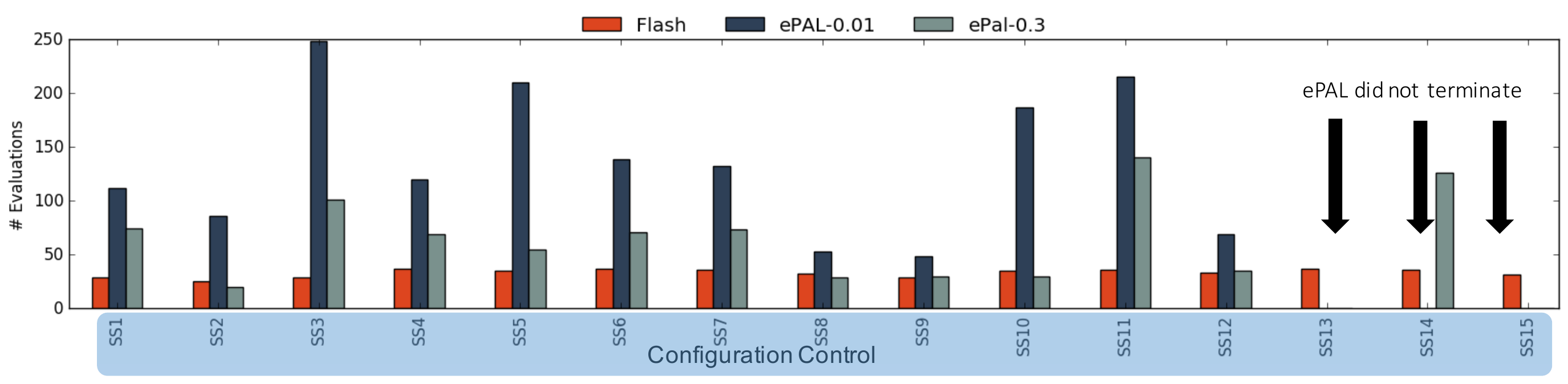}

    \includegraphics[width=\linewidth, height=4cm]{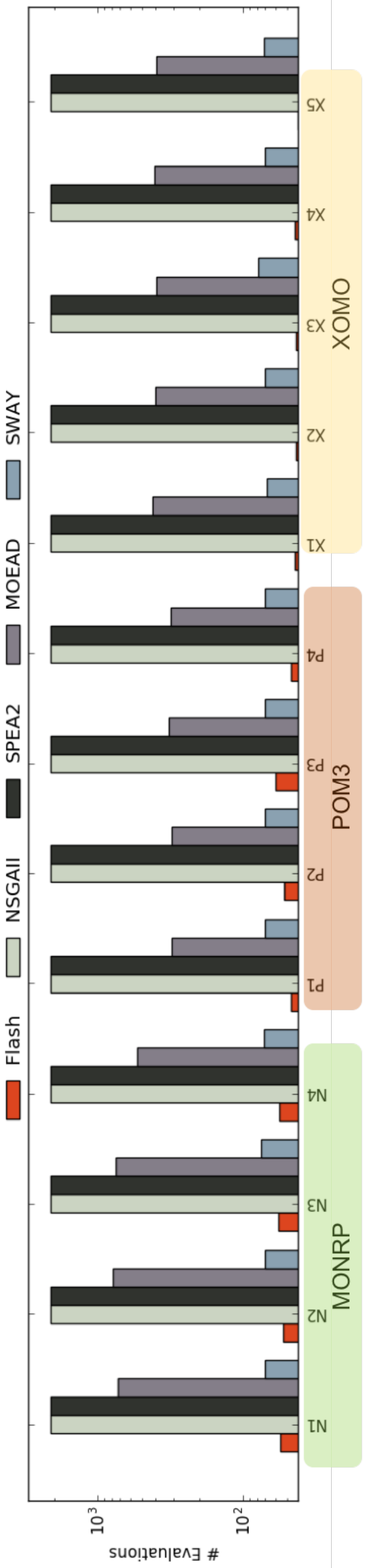}
    \caption{Number of evaluations required by different optimizers to approximate the PF. The {\em y}-axis represents the number of evaluation and {\em x}-axis represents different case-studies. The markers \protect\cusconfig{}, \protect\cusmonrp{}, \protect\cuspom{} and \protect\cusxomo{} represent configuration control, MONRP, POM and XOMO models respectively.} \label{fig:fast_opti}
\end{figure*}

XOMO combines four waterfall software process models developed by Boehm et. al~\cite{me09e}. The inputs to XOMO are the project descriptors, which can be changed by a management decision.  For example, if a manager wants to 
 (a)~{\em relax schedule pressure}, they set {\em sced} to its minimal value;
(b)~to {\em reduce functionality}  they 
halve the value of {\em kloc} and reduce the size of the project
database (by setting {\em data=2});
(c)~to {\em reduce quality} (in order to race something
to market) they might move to the lowest reliability, minimize the documentation work, the complexity of the code, and
reduce the schedule pressure to some middle value. In the language 
of XOMO, this last change would be {\em rely=1, docu=1, time=3, cplx=1}.red
The optimization goals considered in this paper are: minimize risk, effort, cost, and time required for a project. We consider 5 variants of the XOMO in this work, which have been taken from NASA's Jet Propulsion Laboratory~\cite{menzies2009avoid}. For more details refer to~\cite{chen2017beyond}.

%\subsection{Measuring ``Comprehensibility''} 

%\section{Experimental Models}
%\subsection{Research Questions}
% In the past, the SBSE community has used EAs to solve multi-objective problem which requires thousands of evaluations. The results of EAs have been very impressive indeed considering that it is a meta-heuristic search and did not involve explicit programming of the optimization process. The only modification that practitioners had to make was to embed the domain knowledge in form of new reproduction operators. EAs cannot be used in cases where the evaluating hundreds to solutions is not feasible.  

% Our research questions are geared towards finding an optimizer which can be used when evaluating hundreds of solution is not feasible. In this paper, we introduce a new approach called FLASH, which is faster, more succinct and effective as other optimizer. 
% Our proposal is to embrace a Bayesian-inspired search process and build a surrogate model which is easy to train and would scale to higher dimensional data. Therefore to asses the feasibility and usefulness of FLASH, we consider the following:
% \begin{enumerate}
% \item The predicted PF is found by using supplementing the search process by a surrogate model is close to actual PF. 
% \item The solutions sampled by FLASH can then be used to provide a comprehensive picture of promising solutions.
% \item The prediction PF can be obtained using fewer evaluation when compared to more expensive techniques such as traditional EAs or is more scalable than ePAL.
% \end{enumerate}

\section{Results}

To comparatively assess FLASH,  we repeated the following procedure twenty times (each time with a
different random number seed). Firstly,  we generate 10,000 random solutions for each model\footnote{.
For MONRP, we keep generating until 10,000 solutions with valid decisions  are found, while for SS* the solution space is fixed}.
Next, we give each optimizer however many solutions it wants; specifically,
NSGA-II, SPEA2, MOEA/D want 100; the rest take 10,000. The algorithms then run till their internal
termination criteria trigger at which point some final {\em best} set of decisions are returned. 

These runs were instrumented to answer three research
questions:
\bi
\item
\textbf{RQ1:} \textit{Is FLASH a fast optimizer?}
\item
\textbf{RQ2:} \textit{Can FLASH generate a comprehensive and a succinct description of the search space?}
\item
\textbf{RQ3:} \textit{Is FLASH as effective as other optimizers?}  
\ei
 
\subsection{RQ1. Is FLASH a fast optimizer?}

Figure~\ref{fig:fast_opti} shows the median 
number of evaluations used by each optimizer while processing our models. In that figure, 
the markers \protect\cusconfig{}, \protect\cusmonrp{}, \protect\cuspom{} and \protect\cusxomo{} represent configuration control, MONRP, POM and XOMO respectively.

Note that the SS13, SS14, SS15 results do not list
all the ePAL versions since some of these did not
terminate (even after ten hours of execution -- a pragmatic choice). The reason for this can be seen in  Figure~\ref{fig:constrained_case_studies}: these models use more than 10 decisions and the Gaussian process models
used by ePAL does not scale beyond 10 dimensional decisions.

The obvious feature of Figure~\ref{fig:fast_opti} is
that FLASH used fewer evaluations that other methods in $\frac{23}{28}$ of the models run here. Further, for some
models (MONRP, POM, XOMO), FLASH terminates
after orders of magnitude fewer evaluations.  Hence,
our answer to {\bf RQ1} is

      \begin{figure*}[!tbh]
    \centering
    \includegraphics[ height=5.8cm]{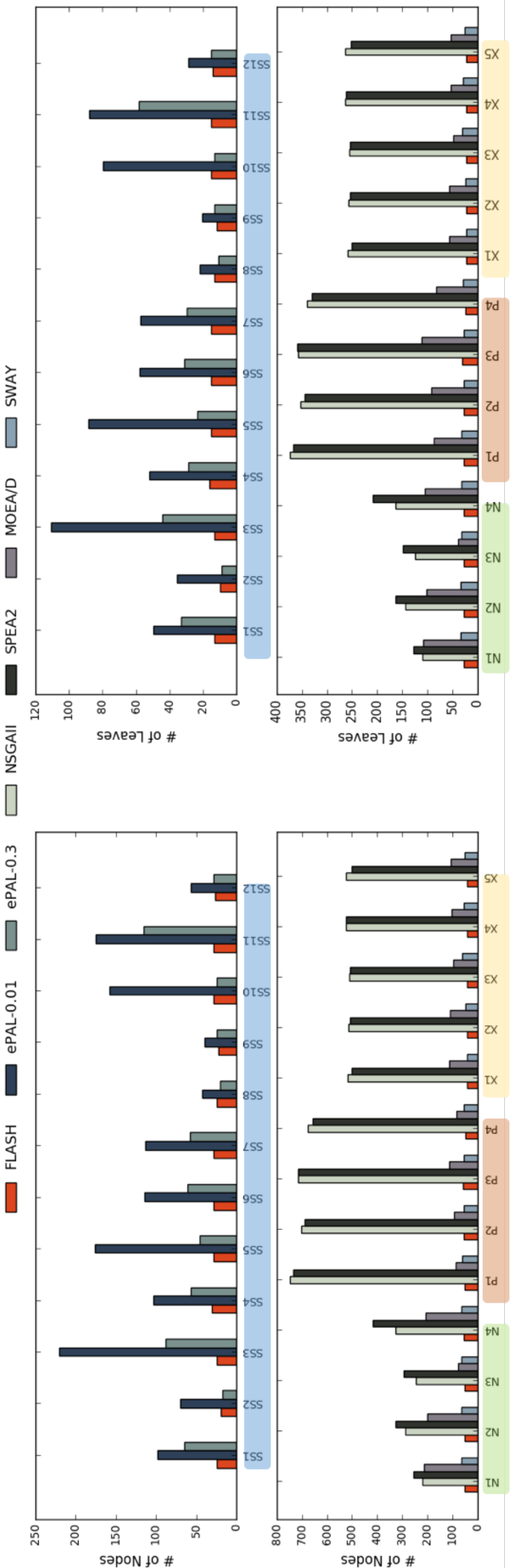}
    \caption{Comparison of comprehensiveness of decision trees (CART) trained using the solutions evaluated by different optimizers. The markers \protect\cusconfig{}, \protect\cusmonrp{}, \protect\cuspom{} and \protect\cusxomo{} represent configuration control, MONRP, POM and XOMO respectively.  A comprehensive tree would have fewer number of nodes as well as leaves. CART trained using points sampled by FLASH has the fewest number of nodes and leaves, which makes it very comprehensive and useful for the decision maker.} \label{fig:succint}
    \end{figure*}

    \begin{figure}[!t]
    \centering
 \includegraphics[width=0.5\textwidth]{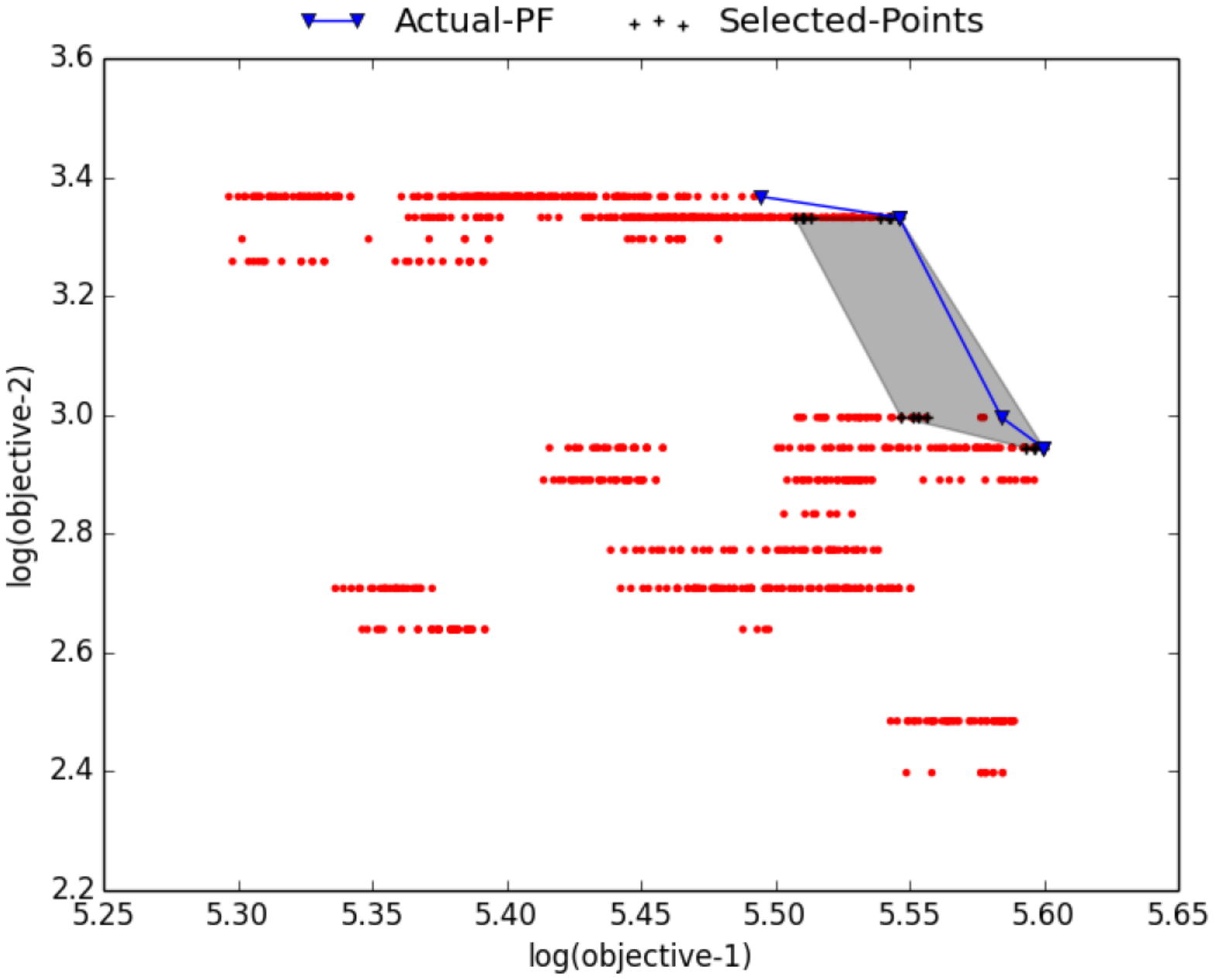}
% \centering
% \fbox{\includegraphics[width=5cm, height=5cm]{Figures/explain_flash}}
%  \begin{lstlisting}[
% escapechar=@,
% basicstyle=\footnotesize\ttfamily,
% keepspaces=true,
% frame=none,
% numbers=left
% ]
% @\bh@sccp=0@\eh@
% |    print_used_types=0
% |    |    ipsccp=0 (6.5)
% |    |    ipsccp=1
% |    |    |    x[10]=0
% |    |    |    |    x[5]=0 (12.5)
% |    |    |    |    x[5]=1 (13)
% |    |    |    x[10]=1
% |    |    |    |    instcombine=0 (18)
% |    |    |    |    instcombine=1 (20.5)
% |    @\bh@print_used_types=1@\eh@
% |    |    ipsccp=0 (14)
% |    |    @\bh@ipsccp=1@\eh@
% |    |    |    time_passes=0 
% |    |    |    |    inline=0 (23)
% |    |    |    |    inline=1 (26)
% |    |    |    @\bh@time_passes=1@\eh@
% |    |    |    |    jump_threading=0 (28)
% |    |    |    |    @\bh@jump_threading=1@\eh@ (31.3)
% sccp=1
% |    ipsccp=0
% |    |    jump_threading=0 (2.5)
% |    |    jump_threading=1 (0.5)
% |    ipsccp=1
% |    |    print_used_types=0 (4.5)
% |    |    print_used_types=1 (9.67)
% \end{lstlisting}
\caption{\small{The examples evaluated
by FLASH when processing the LLVM model.
The blue line connects nodes in the Pareto frontier.
The graph region shows the examples selected
by the branch marked in gray in Figure~\ref{fig:dom1}.
Note that that branch selects for nearly the entire frontier.}}
\label{fig:flash_DT} 
\end{figure}

\vskip 1ex
 \begin{myshadowbox}
        FLASH evaluates fewer points than the ePAL and EAs. The number of evaluations used by FLASH is an order of magnitude less that the traditional EAs and uses only half the number of evaluations as ePAL-0.3.
 \end{myshadowbox}

\begin{figure*}[tbh]
    \includegraphics[width=\linewidth, height=9cm]{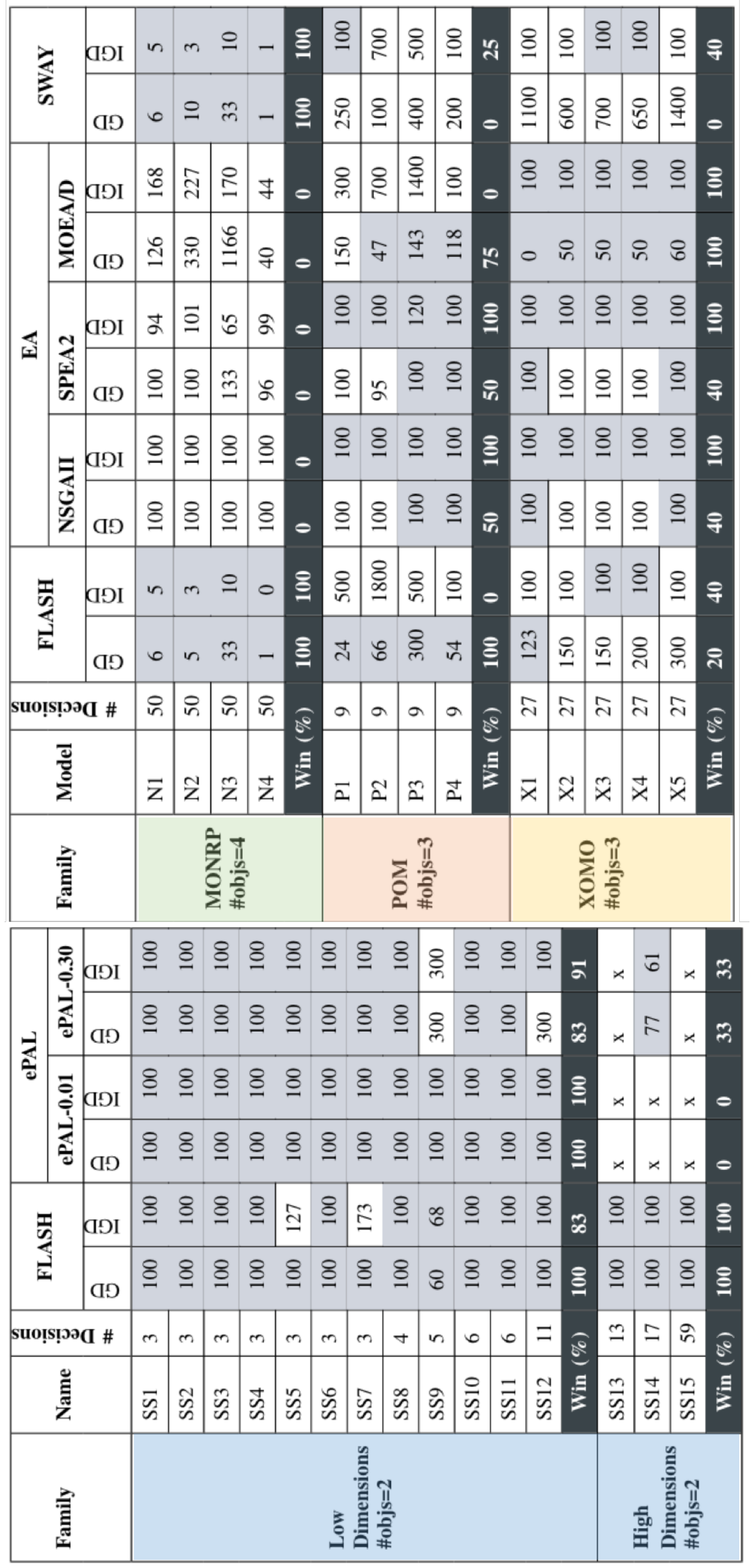}
    \caption{Statistical comparisons of FLASH,
    ePAL, NSGA-II, SPEA2, MOEA/D, SWAY.
    Performance measures are GD (Generational Distance)
    and IGD (Inverted Generational Distance). For these measures
    {\em less} is {\em better};
    ``x'' denotes cases where algorithms did not terminate within a reasonable amount of time (10hrs). 
    All numbers are expressed as percentages
    of the ePAL-0.01 results (left-hand-side) and NSGA-II results (right-hand-side). That is ``100'' means ``as good
    as the baseline methods''. 
    Gray cells denote results from a statistical analysis
    (the Scott-Knot effect size test). Any cell in gray is as 
    ``as good as the best'' in any row. Cells with a black background summarize how often a model was ``as good as the best'' within a group of models.
     } \label{fig:effectiveness}
\end{figure*}

\subsection{RQ2. Can FLASH generate a comprehensive and a succinct description of the space?}
As discussed above, an easy way to understand the search space is to build a domination tree using the points sampled by a optimizer. Since, a decision maker (human) needs to go through all the branches of the tree it is important that the tree is small.
In this paper, we measure comprehensiveness of the optimizer by two measures: (i) number of branch points or nodes and (ii) number of leaves in the decision tree build using the solutions sampled by the optimizers.  
Figure~\ref{fig:succint} shows the number of nodes and  leaves build on the examples
evaluated  by each optimizer for different models. The {\em y}-axis (left) measures the number of nodes in the tree whereas the {\em y}-axis (right) measures the number of leaves in the tree. In this figure:

\begin{itemize}

    \item The optimizer which results in the lesser number of nodes is preferred over the ones with large number of nodes.

    \item Leaves represent the granularity of values (domination scores). Lower number of leaves is preferred over larger number of leaves since such domination trees are easier to analyze.
    
\end{itemize}

Overall in Figure~\ref{fig:succint},  we see that FLASH generates domination trees which have both fewer nodes and fewer leaves, which is orders of magnitude less that trees generated by traditional EAs. For techniques like ePAL and SWAY the difference between the trees size is less magnified that those of other techniques. 

It reasonable  to ask whether  domination
trees generated in this way
are of of any use to the decision maker. 
To answer that question, we refer back to 
 Figure~\ref{fig:dom1},
and also   Figure~\ref{fig:flash_DT}.
Recall that Figure~\ref{fig:dom1} shows a summary
of the examples evaluated by FLASH as it 
   optimized the  LLVM configurations.
That tree had a branch marked in gray-- this was the branch leading to the examples that dominated most other examples.

Figure~\ref{fig:flash_DT} shows in black the examples that are selected by the highlighted tree branch of Figure~\ref{fig:flash_DT}. The graph polygon of that figure
shows the region covered by those examples.
Note that the gray region covers nearly all the Pareto frontier (shown as a blue line). 

To summarize, examples sampled by FLASH is much lower than the other optimizer. Further, these few samples can generate
very succinct domination trees that can be used to
easily read a description of an approximation to
the Pareto frontier. Hence we say:
\vskip 1ex
 \begin{myshadowbox}
        The points sampled by FLASH is useful to build smaller trees, which provides a comprehensive view of the search space. 
 \end{myshadowbox}

\subsection{RQ3. Is FLASH as effective as other optimizers?}

Figure~\ref{fig:effectiveness} shows a statistical analysis using the Scott-Knott effect size test. In this figure,
cells marked with an ``x'' show where some optimizer failed to terminate. Note that ePAL rarely terminated for SS13, SS14 and SS15.

In Figure~\ref{fig:effectiveness}, 
 the median values of 
 Generational Distance (GD) and 
Inverted Generational Distance (IGD)  are shown. These values are collected over 20 runs and are generated
by different optimizers for various case studies.
All these numbers are expressed as ratios of the state-of-the-art result known prior to this paper. Hence:
\bi
\item
A value of 100 means ``same as the prior state-of-the-art'';
\item Values less than 100 can indicate an improvement;
\item Values more than 100 can indicate an optimizer
performing worse than the state-of-the-art.
\ei
To define those ratios, we used 
ePAL-0.01, NSGA-II as the state-of-the-art for configuration control and other models. For SS13-15, FLASH has been used as the state-of-the-art since ePAL-0.01 did  not terminate for these systems.

Note one quirk of  Figure~\ref{fig:effectiveness}: it has many cells with values of ``100''. There are two reasons why this so,
Firstly, many of these columns are ``100'' by definition.
Recall that we use ePAL-0.01 and NSGA-II as the reference optimizers to define ``100''. Hence, all values in those columns will be 100.
Secondly,
many of the configuration problems have only
a few examples of their Pareto frontier; e.g. see the four blue points in Figure~\ref{fig:flash_DT} that define the LLMV Pareto frontier. When the target is that simple, many optimizers will find the same solutions and, hence, score 100 within our ratio calculations.

As to the gray cells  Figure~\ref{fig:effectiveness},
these denote results that are statistically insignificantly different from the best result for any row.
That is, any cell in gray is ``as good as the best''.

One way to get a quick summary of this table is to
read the black cells. These cells
count the number of times an optimizer
was marked as ``as good as the best''.
Looking over those black summary cells:
\bi
\item
For the SS* models:
\bi
\item FLASH  clearly wins over
ePAL;
\item This is particularly true for the configuration problems
with most decisions such as SS13, SS14, SS15.
For those problems, ePAL rarely terminated while
FLASH always did.
\ei
\item
For the  MONRP models:
\bi
\item
FLASH ties with SWAY for best place;
\ei
\item
For the POM models:
\bi
\item
FLASH performs  well, measured in terms of Generational Distance (GD).
\item
But when looking at Inverted Generational Distance,  
NSGA-II and SPEA2 perform best.
\ei
\item
For the XOMO models:
\bi
\item
FLASH and SWAY are clear losers.
\ei
\ei

Clearly FLASH is always not the best optimizer-- and we 
should not expect it to be.
Wolpert and Macready~\cite{wolpert1997} showed
that we can never expect that any optimizer always works
well for all problems\footnote{Specifically:
for any optimization algorithm, any elevated performance over one class of problems is offset by performance over another class.}.
Still, there is much here to recommend FLASH:
\bi
\item
FLASH can handle models
with many more decisions that ePAL (see the SS13, SS14, SS15 results). For those configuration problems, ePAL rarely worked and FLASH always worked.
\item
Across all of Figure~\ref{fig:effectiveness}, there are more successes with FLASH than any other approach.
\item
For some real-world problems, it might be indeed useful to use an optimizer which terminates faster and provide a comprehensible report of the search space rather than producing better performance after a long and expensive optimization process.
\ei
That said, there might be a systematic reason why FLASH
is failing for models like XOMO and not perform best for models
like POM. As noted in
Figure~\ref{fig:unconstrained_case_studies}, XOMO and POM
are {\em unconstrained models}-- which means that good solutions can be spread across large regions of the decision space.  
That is,   we conjecture that for models with several
objectives and no constraints (e.g. POM and XOMO),
FLASH might need to be augmented with some solution
diversity operator.

To summarize, FLASH is very effective on problems for constrained problems. However, the basic algorithm might need
more work for unconstrained problems with objectives greater than 2.  
\vskip 1ex
 \begin{myshadowbox}
        FLASH is effective for configuration control problem and constrained model like MONRP. However, more work is required for   unconstrained problem with dimensions more than 2. 
 \end{myshadowbox}

\section{Discussion}

\subsection{What is the trade-off between the number of lives and the number of measurements?}
FLASH requires that the practitioner defines a termination criterion (\textit{lives} in our setting) before the optimization process commences. The termination criterion preempts the process of searching when the sampled points does not add to the already seen PF. In our experiments, the number of evaluations depends on the termination criterion (\textit{lives}). An early termination of the FLASH would lead to sub-optimal solutions, while late termination would result in resource wastage. We performed a careful manual tuning of the stopping criterion and found empirically that the \textit{lives}=10 is a sweet spot between achieving lower number of evaluations and finding the actual PF.
However, in real-world scenarios, budget allocated for optimization would influence `\textit{lives}'.

\subsection{Why Decision Trees is used as the surrogate model?}
Decision Trees is a very simple way to learn rules from a set of examples and can be viewed as a tool for the analysis of a large dataset. The reason why we chose CART is two fold. Firstly, it is shown to be \textit{scalable} and there is  a growing interest to find new ways to speed up the decision tree learning process~\cite{su2006fast}. 
Secondly, decision tree is able to \textit{describe} with the tree structure the dependence between the decisions and the objectives, which is useful for induction.
These are primary reason for choosing decision-trees to replace Gaussian Process as the surrogate model for FLASH.

% % \subsection{Can rules learned by Decision Tree used to guide the process of search?}
% % Currently FLASH does not explicitly reflect on the decision tree to choose the next point to select. But, rules learned by Decision Tree can definitely be used to guide the process of searching. Though we have not tested this approach, we hypothesize that if a decision tree is trained using indicator domination scores, then the rules-branches leading to leaf with highest mean domination score can be used to guide selection of next point. We leave this for a future work. 

% % \subsection{Why is FLASH less effective in unconstrained problem?}
% % There are multiple reasons for ineffectiveness of FLASH.
% % Firstly, unlike traditional EAs which explores 20,000 (in our experiments), FLASH starts with 10,000 valid solutions. 
% %     Secondly, as the number of dimensions grow the points required to represent the PF grows exponentially. Since, FLASH just evaluates at an average 60 solutions for the POM and XOMO, when compared to 100 non-dominated points found by traditional EAs.
% %   Thirdly,  FLASH currently does not have any strategy to preserve diversity and primarily depends on the $\Omega_I$ score of individuals to decide which point to sample. New strategies can be very useful to encourage spread and convergence, which is left for the future work.

\subsection{Why FLASH and not SWAY?}
SWAY is a novel technique which uses over-sampling as well as approximated principal component to sample points. However, approximation of the the principal component is particularly challenging since, it require careful selection of the distance function used during its computation. This has been described by the authors and mentions how embedding domain knowledge into the search process can be challenging~\cite{chen2017beyond}. However, FLASH makes no such assumption since CART is agnostic distribution of the decision values. This quality makes FLASH a off-the-shelf optimizer which can be used without embedding any domain knowledge.

\section{Threats to Validity}
{\em Reliability} refers to the consistency of the results obtained
from the research.  For e.g.,   how well can independent researchers
could reproduce the study? To increase external
reliability, we took care to either  clearly define our
algorithms or use implementations from the public domain
(SciKitLearn)~\cite{scikit-learn}. All code used in this work are available
on-line.

{\em Validity} refers to the extent to which a piece of research
investigates what the researcher purports to investigate~\cite{SSA15}.
{\em Internal validity} concerns with whether the differences found in
the treatments can be ascribed to the treatments under study. 

For the case-studies
relating to configuration control, we cannot measure all possible configurations in reasonable time. Hence, we sampled only few hundred configurations to compare prediction to actual values. We are aware that this evaluation leaves room for outliers and that \textit{measurement bias} can cause false interpretations~\cite{me12d}. We also limit our attention to predicting PF for a given workload, we did not vary benchmarks.

\textit{Internal bias} originates from the stochastic natural of multi-objective
optimization algorithms. The evolutionary process required many
random operations, same as the FLASH was introduced in this paper.
To mitigate these threats, we repeated our experiments for 20 runs
and report the median of the indicators. We also employed
statistical tests to check the significance in the achieved results.

It is very difficult to find the representatives sample test cases to covers all kinds of domain. We just selected four most
common types of decision space to discuss the FLASH basing on
them. In the future, we also need to explore more types of SBSE
problems, especially the problem with other types of decisions. We aimed to increase {\em external validity} by choosing case-studies from different domains.

\section{Conclusion}
Traditionally EAs have been used to solve various SBSE problems, but they requires a large number of evaluations to find a set of non-dominated solutions. This might not be useful in situations where evaluating a single solution is expensive,
or users want some succinct justification for why the model is telling them to do this, or that.

Meanwhile, in the machine learning community, Bayesian Optimization is traditionally used for parameter tuning of machine learning algorithms, but it does not scale beyond a moderate number of dimensions. 

To overcome these shortcomings of EAs and BO, we introduce FLASH -- a hybrid algorithm which  is \textit{fast} in terms of evaluation, \textit{scalable} when compared to BO and \textit{effective} for constrained models. The solutions sampled by FLASH during the process of optimization can be used to build \textit{small and comprehensible} trees which helps in analysis search space. 

To compare FLASH with various state-of-the-art optimizers (ePAL, NSGA-II, SPEA2, MOEA/D and SWAY), we conducted a number of experiments on 15 real-world configurable system as well as on 13 variants of 3 different case studies to demonstrate the qualities of FLASH. We observed that FLASH is effective to find the points very close to the Pareto frontier  for multi-objective constrained problems using fewer evaluations than the state-of-the-art optimizers.

In terms of future work, the two clear directions for this research are case studies with more models and  newer     operators to increase the diversity
of solutions found in FLASH.

To conclude, we make two observations. Firstly, EAs is definitely not a silver bullet to solve all types of problems in SBSE. There are various ways to solve a problem and for problems where evaluations is resource intensive, alternative techniques such as FLASH can be very effective.  Secondly, different communities tackle similar problems and many not be fully aware
of  advances in other communities. This paper experiments with ideas from fields of machine learning, SBSE and software analytics to create FLASH. which is a fast, comprehensible, scalable and effective optimizer.
We hope this paper inspires other researchers to look further afield than their home discipline.

\balance
%\vspace*{0.5mm}
\bibliographystyle{ACM-Reference-Format}

% \balance

\end{document}